\documentclass[aps,prb,reprint,superscriptaddress,amsmath,amssymb,showpacs]{revtex4-1}%

\renewcommand{\AA}{\mathcal{A}}

\newcommand{\CC}{\mathcal{C}}
\newcommand{\LL}{\mathcal{L}}

\newcommand{\dd}{\mathrm{d}}

\newcommand{\ii}{\text{i}}

\newcommand{\av}[1]{\langle #1 \rangle}
\newcommand{\avg}[2]{\langle #1 \rangle_{#2}}
\newcommand{\expo}[1]{\text{exp}\left( #1 \right)}

\newcommand{\tS}{\text{S}}
\newcommand{\tL}{\text{L}}
\newcommand{\tR}{\text{R}}

\newcommand{\tB}{\text{B}}
\newcommand{\tSB}{\text{SB}}

\newcommand{\e}{\text{e}}

\newcommand{\unit}[1]{\,\mathrm{#1}}
\renewcommand{\vec}[1]{\mathbf{#1}}

\usepackage{braket}
\usepackage{subfigure}
\usepackage{graphicx}
\usepackage{xcolor}
\usepackage{dsfont}
\usepackage{pstricks}

\graphicspath{{./overview_5e-2g_06lo/}{../../../HEOM/programs/el_vib_bath/adiabatic_regime/Plots/}{../../../HEOM/programs/system_vib_pola_trafo/Rainer_PHD_model/Plots/}{../../../HEOM/programs/Pade_decomposition/Results/}{./Plots_mm_F_V/}{./Plots_1m_fig/}{./Plots_BDBT/}}

\begin{document}


\title{Hierarchical Quantum Master Equation Approach to  Electronic-Vibrational Coupling in Nonequilibrium Transport through Nanosystems}

\author{C.\ Schinabeck}
\affiliation{Institut f\"ur Theoretische Physik und Interdisziplin\"ares Zentrum f\"ur Molekulare
Materialien, \\
Friedrich-Alexander-Universit\"at Erlangen-N\"urnberg,\\
Staudtstr.\, 7/B2, D-91058 Erlangen, Germany
}
\author{A.\ Erpenbeck}
\affiliation{Institut f\"ur Theoretische Physik und Interdisziplin\"ares Zentrum f\"ur Molekulare
Materialien, \\
Friedrich-Alexander-Universit\"at Erlangen-N\"urnberg,\\
Staudtstr.\, 7/B2, D-91058 Erlangen, Germany
}
\author{R.\ H\"artle}
\affiliation{Institut f\"ur Theoretische Physik, Friedrich-Hund-Platz 1, D-37077 G\"ottingen, Germany}
%
%
\author{M.\ Thoss}
\affiliation{Institut f\"ur Theoretische Physik und Interdisziplin\"ares Zentrum f\"ur Molekulare
Materialien, \\
Friedrich-Alexander-Universit\"at Erlangen-N\"urnberg,\\
Staudtstr.\, 7/B2, D-91058 Erlangen, Germany
}



\date{\today}

\begin{abstract}
Within the hierarchical quantum master equation (HQME) framework, an approach is presented, which allows a numerically exact description of nonequilibrium charge transport in nanosystems with strong electronic-vibrational coupling. The method is applied to a generic model of vibrationally coupled transport considering a broad spectrum of parameters ranging from the nonadiabatic to the adiabatic regime and including both resonant and off-resonant transport. We show that nonequilibrium effects are important in all these regimes. In particular in the off-resonant transport regime, the inelastic co-tunneling signal is analyzed for a vibrational mode in full nonequilibrium, revealing a complex interplay of different transport processes and deviations from the commonly used \emph{$G_0/2$-thumb-rule}. In addition, the HQME-approach is used to benchmark approximate master equation and nonequilibrium Green's function methods. 

\end{abstract}

\pacs{}

\maketitle

%
Nanosystems are often characterized by strong coupling between electronic and vibrational or structural degrees of freedom.
Examples include 
single-molecule junctions,\cite{Tal2008,Secker2011,Neel2011,Lau2016} nanoelectromechanical systems\cite{Craighead2000,Ekinci2005} as well as suspended carbon nanotubes.\cite{Weig2004,Sapmaz2006,Leturcq2009}
Strong electronic-vibrational coupling manifests itself in vibronic structures in the transport
characteristics and may result in a multitude of 
nonequilibrium
phenomena such as current-induced local heating and cooling,
multistability, switching and hysteresis,
as well as decoherence,  which have been observed experimentally
\cite{Gaudioso2000,Pop2005,Molen2010,Ballmann2012}
and have been the focus of theoretical studies.\cite{Galperin2005,Leijnse2008,Haertle2011,Wilner2013,Erpenbeck2015} 
While in certain parameter regimes, approximate methods based on, e.g., scattering theory, master equations or nonequilibrium Green's functions (NEGF) have provided
profound physical insight into transport mechanisms,\cite{Ness2001,Mitra2004,Cizek2004,Galperin2005,Koch2005,Galperin2006,Frederiksen2007,Leijnse2008,Haertle2011,Wilner2013,Schinabeck2014,Erpenbeck2015} the theoretical study of strong coupling situations often requires the application of methods that can be systematically converged, i.e.\ numerically exact methods.
Methods developed in this context include path integral approaches,\cite{Muehlbacher2008,Schiro2009,Huetzen2012,Simine2013} the scattering state numerical renormalization group technique\cite{Jovchev2013} and the multilayer multiconfiguration time-dependent Hartree method. \cite{HWang2009,HWang2013,Wilner2013,Wilner2014}

In this paper, the hierarchical quantum master equation (HQME) approach is formulated to study nonequilibrium transport in systems with strong electronic-vibrational coupling.
The HQME approach generalizes perturbative master equation methods by including higher-order contributions as well as non-Markovian memory 
and allows for the systematic convergence of the results. This approach was originally developed by Tanimura and Kubo in the context of relaxation dynamics.\cite{Tanimura1989,Tanimura2006}
Yan and coworkers\cite{Jin2008,Zheng2013} as well as H\"artle et al.\cite{Haertle2013a,Haertle2015} have used it to study charge transport in models with electron-electron interaction.
An approximate formulation of the HQME method for the treatment of electronic-vibrational coupling was recently proposed.\cite{Jiang2012} Here, we apply the HQME methodology 
for the first time within a numerically exact formulation to treat nonequilibrium transport in nanosystems with strong electronic-vibrational coupling. In contrast to other numerically exact approaches, the HQME method is directly applicable to steady state transport without time propagation, which is an advantage for systems with slow relaxation.

We apply the methodology to study transport phenomena in a broad range of parameters including off-resonant and resonant transport as well as the adiabatic and nonadiabatic transport regimes. 
In the off-resonant transport regime, it is shown that the peak-dip transition of the first inelastic cotunneling feature does not follow the commonly used $G_0/2$-thumb-rule,\cite{Paulsson2005,Vega2006,Tal2008,Avriller2009,cuevasscheer2010} if the nonequilibrium excitation of the vibration is taken into account.
The HQME method is also applied to benchmark approximate master equation and NEGF methods.
To be specific, we adopt in the following the terminology used in the context of quantum transport in molecular nanojunctions. It should be noted, though, that the methodology is applicable also to other nanosystems with strong electronic-vibrational coupling as mentioned above.

We consider a generic model of vibrationally coupled electron transport in molecular junctions with the Hamiltonian (we use units where $\hbar=e=1$)
\begin{align*}
 H = & \epsilon_0 d^{\dagger} d + \sum_{k \in \tL/\tR} \epsilon_{k} c_{k}^{\dagger} c_{k} + \sum_{k \in \tL/\tR} (V_{k} c_{k}^{\dagger} d + V^*_{k} d^{\dagger} c_{k}) \\
  &+ \Omega a^{\dagger} a + \lambda (a +a^\dagger) d^{\dagger} d.
\end{align*}
A single electronic state with energy $\epsilon_0$ located on the molecular bridge is coupled to a continuum of electronic states with energies $\epsilon_k$ in the macroscopic leads via interaction matrix elements $V_{k}$. The operators $d^\dagger/d$ and $c_{k}^\dagger/c_{k}$ denote the corresponding creation/annihilation operators. We consider a single vibrational mode with frequency $\Omega$, creation/annihilation operators $a^\dagger/a$ and electronic-vibrational coupling strength $\lambda$.
The interaction between the molecule and the left/right lead is characterized by the spectral densities $\Gamma_{\tL/\tR} (\omega)=2 \pi \sum_{k \in \tL/\tR} | V_k |^2 \delta (\omega-\epsilon_k)$. 

To derive the HQME for electronic-vibrational coupling, it is expedient to employing a small polaron transformation, $\tilde H = S H S^\dagger$ with $S=\expo{d^\dagger d (\lambda / \Omega ) \left(a^\dagger -  a \right)}$. Introducing, furthermore, a system-bath partitioning, we obtain $\tilde H= \tilde H_\tS + \tilde H_\tB + \tilde H_\tSB$ with $\tilde H_S=\tilde \epsilon_0 d^{\dagger} d + \Omega a^{\dagger} a$, $\tilde H_\tSB = \sum_{k \in \tL/\tR} (V_{k} X c_{k}^{\dagger} d + \text{h.c.})$ and $\tilde H_\tB=\sum_{k \in \tL/\tR} \epsilon_{k} c_{k}^{\dagger} c_{k}$. 
Thereby, the energy of the electronic state is renormalized by the reorganization energy $\tilde \epsilon_0 = \epsilon_0 - \lambda^2 / \Omega$ and the molecule lead coupling term is dressed by the shift operator $X=\exp\{(\lambda/\Omega)(a-a^\dagger)\}$.

%

As the bath coupling operators $f^\sigma_K (t) = \expo{\ii \tilde H_\tB t} \left( \sum_{k \in K} V_{k} c^\sigma_{k} \right) \expo{-\ii \tilde H_\tB t}$ with $c_k^{-(+)}\equiv c_k^{(\dagger)}$ obey Gaussian statistics,
all information about the system-bath coupling is encoded in the two-time correlation function of the free bath $C^\sigma_K(t-\tau)=\avg{ f_K^\sigma(t) f_K^{\bar \sigma} (\tau) }{\tB}$ with the lead-index $K \in\{\tL,\tR\}$, $\sigma=\pm$ and $\bar \sigma \equiv -\sigma$. To derive a closed set of equations of motion within the HQME method, $C^\sigma_K(t)$ is expressed by a sum over exponentials, $C^\sigma_K(t)=\sum_{l=0}^{l_\text{max}} \eta_{K,l} \e^{-\gamma_{K,\sigma,l} t}$.\cite{Jin2008} To this end, the Fermi distribution is represented by a sum-over-poles scheme employing a Pade decomposition\cite{Ozaki2007} and the spectral density of the leads is assumed as a single Lorentzian $\Gamma_K (\omega)=\frac{\Gamma W^2}{(\omega-\mu_K)^2 +W^2}$, where $\Gamma=\Gamma_\tL=\Gamma_\tR$ denotes the overall molecule-lead coupling strength for a symmetric junction, $\mu_K$ the chemical potential of lead $K$ and $W$ the width of the band. Choosing the latter as $W=10^4 \unit{eV}$, the leads are effectively described in the wide-band limit. A symmetric drop of the bias voltage at the contacts is used.

Following a similar derivation as for a noninteracting model,\cite{Jin2008} the HQMEs for vibrationally coupled transport are obtained as
\begin{align}
\dot \rho^{(n)}_\vec{j} (t) =&- \left( \ii \tilde \LL_\tS + \sum_{i=1}^n \gamma_{j_i} \right) \rho^{(n)}_\vec{j} (t) - \ii \sum_j \tilde \AA^{\bar \sigma} \rho^{(n+1)}_{j,\vec{j}} (t) \nonumber\\
&- \ii \sum_{k=1}^n (-)^{n-k} \tilde \CC_{j_k} \rho^{(n-1)}_{\vec{j} \smallsetminus j_k} (t),
\label{eq:EOM}
\end{align}
with the vector notation $\vec{j}=(j_n, \dots, j_1)$ and multi-index $j=(K,\sigma,l)$.
Thereby, $\rho^{(0)}$ denotes the reduced density operator of the system and $\rho^{(n)}_\vec{j}$ $(n>0)$ auxiliary density operators, which describe bath-related observables such as, e.g., the current $\av{I_K (t)}= \ii \text{Tr}_\tS \left\{ dX \rho_{K,+,l}^{(1)}(t) - 
\text{h.c.}
\right\}$.
The equations differ from those of the noninteracting model by the superoperators $\tilde \AA$ and $\tilde \CC$, which are dressed by the shift operator $X$ and read
\begin{subequations}
\begin{align}
 \tilde \AA^{\bar \sigma} \rho^{(n)} =& d^{\bar \sigma} X^{\bar \sigma} \rho^{(n)} + (-)^n \rho^{(n)} d^{\bar \sigma} X^{\bar \sigma}, \\
\tilde \CC_j \rho^{(n)}  =& \eta^\sigma_{K,l} d^\sigma  X^\sigma   \rho^{(n)} - (-)^n \eta^{\bar \sigma,*} _{K,l} \rho^{(n)} d^\sigma X^\sigma.
\end{align}
\end{subequations}
In the calculations presented below, the coupled set of equations is solved directly for the steady state by setting $\dot \rho^{(n)}_\vec{j} (t=\infty)=0$ ($n \geq 0$). The hierarchy is truncated at a maximum level $n_\text{max}=4$, which provides quantitatively converged results for the electrical current.

While the approach introduced above keeps the vibrational mode as part of the system and thus allows a numerically exact treatment, the approximate HQME approach by Jiang \emph{et al.}\cite{Jiang2012} treats it as part of the bath. As a result of the polaron transformation, 
the modified bath-coupling operators do not obey Gaussian statistics. Consequently, a HQME treatment based on the two-time correlation function neglects nonequilibrium vibrational excitation and partially electronic-vibrational correlations.\cite{Jiang2012} This will be demonstrated below.


%
In the following, we illustrate the performance of the method by applications to representative models covering a broad range of parameters (see Tab.\ \ref{tab:models}). We also use the numerically exact HQME approach to benchmark often used approximate methods including a Born-Markov master equation (BMME),\cite{Mitra2004,Volkovich2011,Haertle2011} a 4th-order $(V_k^4)$ non-Markovian ME\cite{Croy2011,Popescu2013} as well as a NEGF approach within the self-consistent Born\cite{Hyldgaard1994,Galperin2004,Erpenbeck2015} and the full self-consistent Born\cite{Mitra2004,Galperin2004,Ryndyk2006,Dash2011} approximation ((F)SCBA). The FSCBA treats both electrons and vibrations self-consistently, whereas the latter are not incorporated self-consistently for the SCBA thus neglecting nonequilibrium vibrational excitation. 

%
\begin{table}[htb!]
 \centering
  \begin{tabular}{c | c c c c c c }
  Model & $ \epsilon_0$ [eV] & $ \tilde \epsilon_0$ [eV] & $\Omega$ [eV] &  $\lambda$ [eV] & $T$ [K] \\ \hline
  1 & 0.3 & 0.228 & 0.2 & 0.12 & 300 \\
  2 & 1.05 & 0.25 & 0.2 & 0.4 & 300 \\
  3 & 0.6 & 0.564 & 0.1 & 0.06 & 100 \\
  4 & 0.6 & 0.528 & 0.2 & 0.12 & 300 
 \end{tabular}
 \caption{Summary of model parameters.}
 \label{tab:models}
\end{table}
\begin{figure}[h!]
	\centering
\includegraphics[width=\columnwidth]{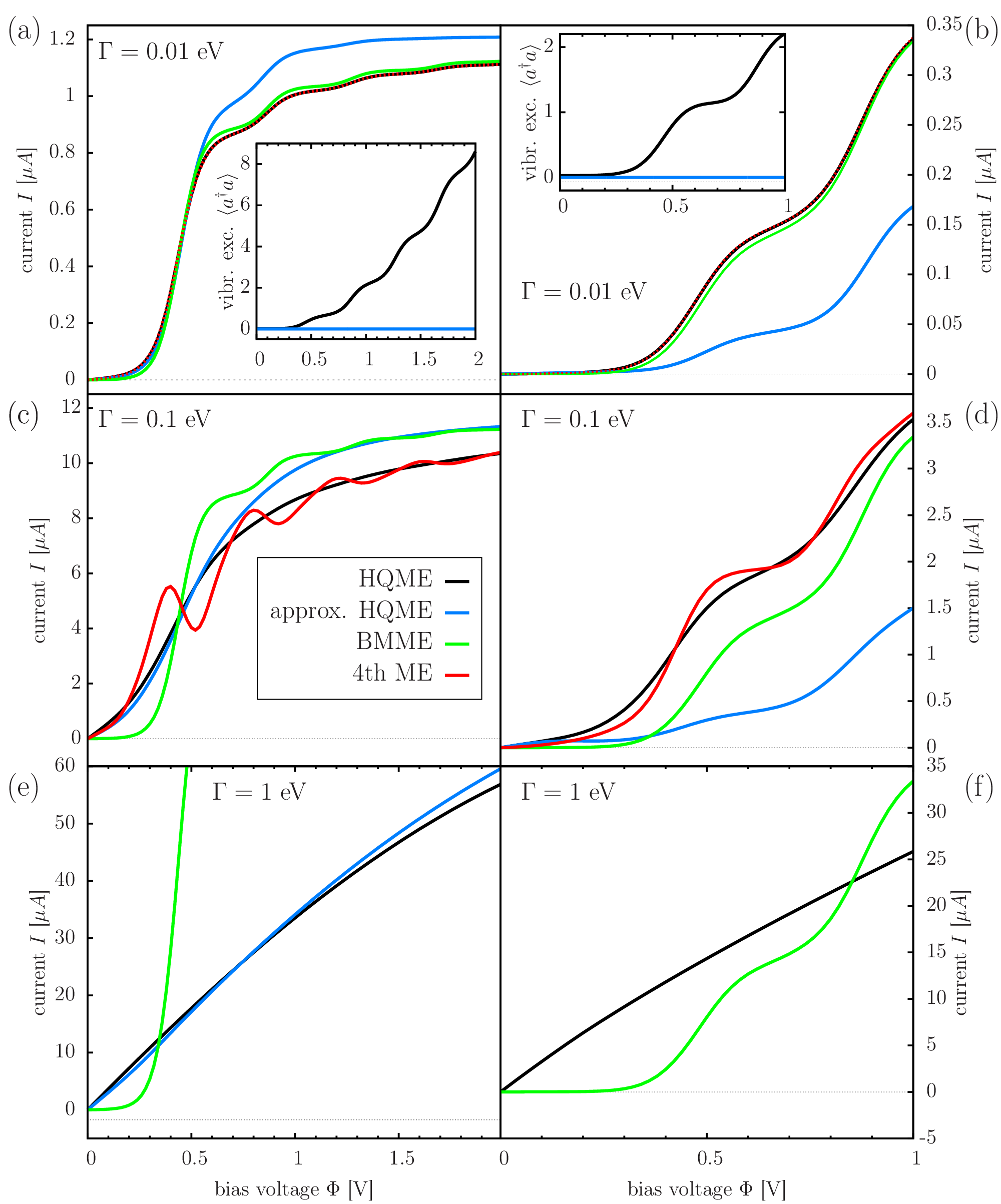}
%
\caption{$I$-$V$s obtained by the accurate HQME approach and different approximate methods. The results are shown for model 1 (a,c,e) as well as model 2 (b,d,f).
		}
\label{fig:model1}
\end{figure}
Fig.\ \ref{fig:model1} shows the current-voltage characteristics ($I$-$V$s) and the average vibrational excitation for moderate ($\lambda/\Omega=0.6$) as well as strong ($\lambda/\Omega=2$) electronic-vibrational coupling  and for a range of molecule-lead coupling strengths $\Gamma$. 
Focussing first on the $I$-$V$s for model 1 ($\lambda/\Omega=0.6$) and $\Gamma=0.01 \unit{eV}$ (Fig.\ \ref{fig:model1}a), corresponding to the nonadiabatic transport regime ($\Gamma
< \Omega$), the accurate HQME results exhibit the typical Franck-Condon step structure. The vibrational excitation depicted in the inset demonstrates the strong nonequilibrium character of the transport process, which results in values significantly larger than the thermal equilibrium value of $4.4 \cdot 10^{-4}$.  The current-induced vibrational excitation results in a suppression of the current for $\lambda/\Omega< 1$.\cite{Haertle2011} As a result, the approximate HQME method of Jiang \emph{et al.},\cite{Jiang2012} which neglects the nonequilibrium vibrational excitation, overestimates the current in the resonant transport regime ($\Phi \gtrsim 2\tilde \epsilon_0$). However, it includes the broadening of the electronic level due to molecule-lead coupling, which is completely neglected in the BMME. The 4th-order ME calculation perfectly agrees with the accurate result in this regime of small molecule-lead coupling.

%

In the regime of strong electronic-vibrational coupling ($\lambda/\Omega=2$, model 2), the first step in the $I$-$V$ (Fig.\ \ref{fig:model1}b) is significantly smaller than for $\lambda/\Omega=0.6$. This is a manifestation of Franck-Condon blockade.\cite{Koch2005} For $\lambda/\Omega > 1$, the transitions between the low-lying vibrational states of the unoccupied and occupied molecular bridge are exponentially suppressed. In this case, the $I$-$V$ obtained by Jiang's approximate HQME approach exhibits a lower current level than the accurate result because the Franck-Condon blockade is more pronounced if the nonequilibrium excitation of the molecular bridge is neglected.\cite{Koch2005}
The 4th-order ME reproduces the accurate result whereas the BMME shows small deviations due to the neglected molecule-lead broadening.


Figs.\ \ref{fig:model1}c,d show $I$-$V$s for moderate molecule-lead interaction, $\Gamma=0.1 \unit{eV}$. The increased molecule-lead interaction results in a broadening of the Franck-Condon steps.
As a result, the deviations of the results obtained by the BMME are more pronounced than for $\Gamma=0.01 \unit{eV}$.
For $\lambda/\Omega=0.6$, the 4th-order ME calculation exhibits spurious oscillations around the accurate result indicating the breakdown of perturbation theory. A similar behavior has already been reported in Ref.\ \onlinecite{Croy2011} for a double quantum dot with Coulomb interaction but without electronic-vibrational coupling.
Remarkably, these oscillations are much less pronounced for $\lambda/\Omega=2$ and $\Gamma=0.1 \unit{eV}$. This can be attributed to the fact that the effective molecule-lead coupling, which determines the range of validity of the perturbative expansion, is given by $|X_\text{max}|^2 \Gamma$.\cite{Eidelstein2013}

For strong molecule-lead coupling ($\Gamma=1 \unit{eV}$),
corresponding to the adiabatic transport regime ($\Gamma > \Omega$), the accurate HQME results predict almost linear $I$-$V$s (Fig.\ \ref{fig:model1}e,f). For moderate electronic-vibrational coupling ($\lambda/\Omega=0.6$), the approximate HQME result shows rather good agreement, indicating negligible vibrational nonequilibrium effects.
For strong molecule-lead coupling, the BM-approximation and the 4th-order ME treatment are invalid. In the case of additional strong electronic-vibrational coupling, also the approximate version of the HQME method fails (data not shown).
\begin{figure}[h!]
	\centering
\includegraphics[width=0.95\columnwidth]{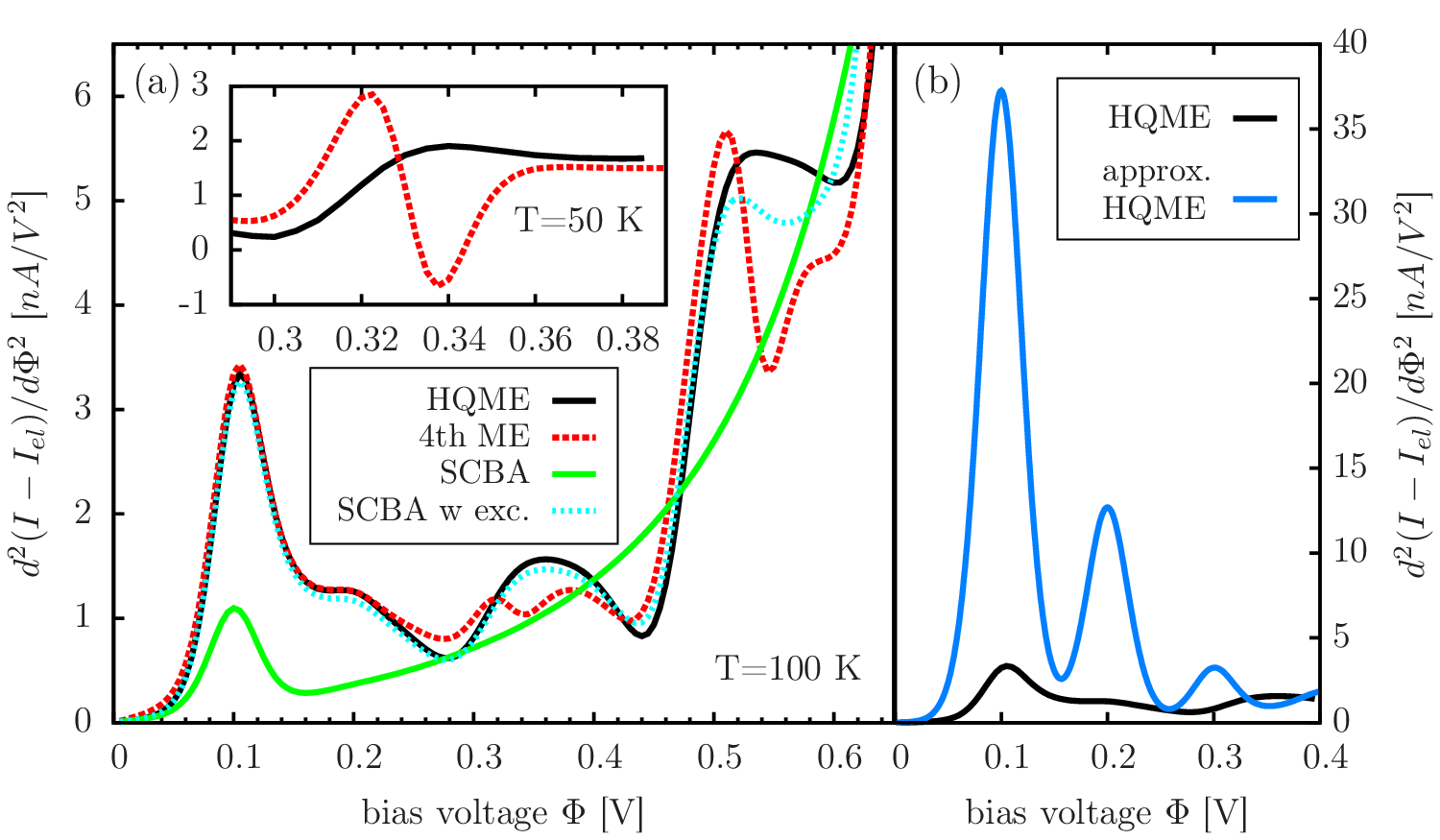}
\caption{IETS for model 3 and $\Gamma=6.667 \cdot 10^{-3} \unit{eV}$. The purely electronic contribution $I_{\rm el}$ has been substracted for a better resolution of inelastic effects. The 4th-order ME as well as the NEGF-SCBA approach are compared to the accurate HQME approach in panel (a). The inset shows the peak-dip structure at $\Phi=0.328 \unit{V}$ for $T=50 \unit{K}$. Panel (b) depicts a comparison with the approximate version of the HQME approach.
}
\label{fig:model2}
\end{figure}
%
%
Next, we consider in more detail the off-resonant transport regime for low bias voltages $\Phi < 2\tilde \epsilon_0$. In this regime transport is governed by elastic and inelastic cotunneling processes.\cite{cuevasscheer2010}  
The latter result in characteristic structures in the inelastic electron tunneling spectrum (IETS), given by the second derivative of the current $d^2I/d\Phi^2$, which have been observed for many molecular junctions.\cite{Jaklevic1966,Wang2004,Tal2008,Song2009}
Even though we consider a single vibrational mode, we already obtain a rather complex IETS, which is depicted for model 3 in Fig.\ \ref{fig:model2}a. The accurate HQME results exhibit a peak at $\Phi=\Omega$, which marks the onset of inelastic cotunneling via the emission of one vibrational quantum. The satellite peak at $\Phi = 2 \Omega$ corresponding to the emission of two vibrational quanta is suppressed and appears as a shoulder because of the overlap with the peak around $\Phi=\Omega$ due to thermal broadening.
For $\Phi \in [0.28,0.44] \unit{V}$, the graph exhibits a structure which results from the superposition of two effects: (i) further inelastic cotunneling peaks at $\Phi=3 \Omega$ and $\Phi=4 \Omega$, the intensity of which is, however, increasingly suppressed and (ii) resonant transport processes facilitated by current induced vibrational excitation. The latter processes include the deexcitation by $n$ vibrational quanta and become active at the thresholds $\Phi= 2 (\tilde \epsilon_0 - n \Omega )$. These resonant transport processes are reflected by peaks in the conductance and thus by a peak-dip feature in the IETS, which is more clearly seen for lower temperature in the inset of Fig.\ \ref{fig:model2}a.

The comparison of the numerically exact HQME results to results of approximate methods for the IETS reveals that the 4th-order ME provides a good approximation for $\Phi \lesssim 0.2 \unit{V}$. For larger voltages it deviates significantly because it misses to some extent the broadening due to molecule-lead coupling. This is especially apparent in the lower temperature result in the inset of Fig.\ \ref{fig:model2}a, which has reduced thermal broadening.
The NEGF-SCBA approach underestimates the height of the first peak at $\Phi =0.1 \unit{V}$ in the IETS by almost 70 \% and essentially misses the second peak around $\Phi =2 \Omega=0.2 \unit{V}$.\cite{Dash2011}  
This deficiency is a consequence of the thermal equilibrium treatment of the vibration. This is demonstrated by the cyan line, which has been obtained by using the average vibrational excitation obtained from the  HQME calculation as input for the SCBA calculation, resulting in good agreement of the IETS with the HQME result for $\Phi \lesssim 0.5 \unit{V}$.
%
The approximate version of the HQME method (solid blue line in Fig.\ \ref{fig:model2}b) overestimates the height of the inelastic cotunneling peaks profoundly. This shows the importance of electronic-vibrational correlations, in particular in the off-resonant transport regime.\cite{HWang2013}

\begin{figure}[h!]
	\centering
%
%
\includegraphics[width=\columnwidth]{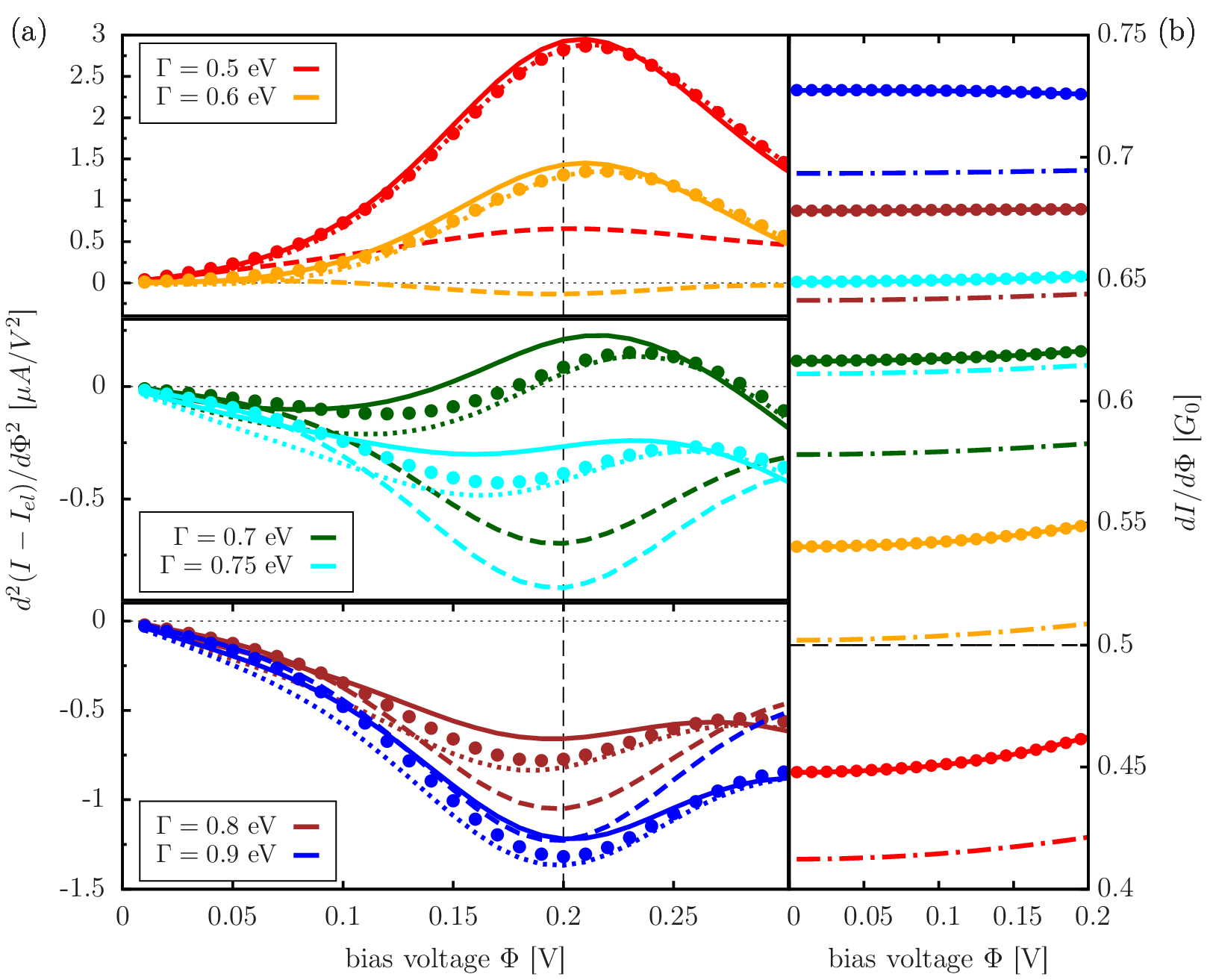}
		\caption{IETS (a) and differential conductance (b) for model 4 and different molecule-lead couplings $\Gamma$.
		The graphs depict HQME results obtained with a truncation after the third (solid lines) and fourth (filled circles) level of the hierarchy. The dashed and dotted lines in panel (a) represent results of a NEGF calculation within SCB- and FSCB-approximation. In panel (b), the conductance-voltage characteristics is also depicted for the non-interacting system by dashed-dotted lines.
		}
\label{fig:model3}
\end{figure}
%

Finally, we consider in  Fig.\ \ref{fig:model3} the change of the IETS line shape upon increase of the molecule-lead coupling, which has been the focus of several theoretical studies recently.\cite{Galperin2004,Galperin2004a,Egger2008,Entin-Wohlman2009} The HQME results show the transition of the inelastic cotunneling feature from a peak ($\Gamma \lesssim 0.6 \unit{eV}$) to a dip ($\Gamma \gtrsim 0.8 \unit{eV}$) via a dip-peak feature in the interval $\Gamma \in [0.7,0.75] \unit{eV}$.
%
Qualitatively, our results do not strictly follow the commonly used \emph{$G_0/2$-thumb-rule},\cite{Paulsson2005,Vega2006,Tal2008,Avriller2009,cuevasscheer2010} which states that  for a system with a zero-bias conductance (determined in the non-interacting case), which is smaller than half of the conductance quantum $G_0$, the IETS exhibits a peak, whereas it shows a dip for higher zero-bias conductance.
This rule was originally derived based on a lowest order perturbative expansion in electronic-vibrational coupling.\cite{Paulsson2005}
Assuming a thermally equilibrated vibration, it was later generalized by Egger\cite{Egger2008} and Entin-Wohlman\cite{Entin-Wohlman2009} \emph{et al.}, who found that the peak-dip transition is not universal at a zero bias conductance of $G_0/2$ but depends on all model parameters. They reported an upper bound of $G_0/2$ for the peak-dip transition.
In contrast, the results obtained for $\Gamma=0.6 \unit{eV}$ (orange circles in Fig.\ \ref{fig:model3}a), corresponding to a zero bias conductance of $0.54\ G_0$ ($0.5\ G_0$ in the non-interacting case) still exhibit a peak in the IETS at $\Phi = \Omega$.
The crossover between the peak- and dip-like structure rather occurs for a zero bias conductance between $0.62\ G_0$ and $0.65\ G_0$ ($0.58\ G_0$ and $0.61\ G_0$ in the non-interacting case) in model 4. This is demonstrated by the green ($\Gamma=0.7 \unit{eV}$) and cyan ($\Gamma=0.75 \unit{eV}$) circles,  which show a dip-peak feature around $\Phi = \Omega$.
Our findings suggest that the deviations from the \emph{$G_0$/2-thumb-rule} result from the nonequilibrium excitation of the vibrational mode. This conjecture is confirmed by the comparison of the HQME results with SCBA- as well as FSCBA-calculations in Fig.\ \ref{fig:model3}.
While the SCBA results, which treat the vibration in equilibrium follow strictly the \emph{$G_0$/2-thumb-rule}, the FSCBA, which incorporates nonequilibrium effects within a perturbative treatment are in rather good agreement with the HQME results.
The comparison of different truncation levels shows that the HQME results for the conductance are quantitatively converged for $n=4$. For the IETS small deviations occur for some of the parameters. This is not very surprising, because the quantity $\dd^2 (I - I_\text{el})/\dd \Phi^2$ is more difficult to converge than the current or the conductance (Fig.\ \ref{fig:model3}b).  

In summary, the HQME method presented here allows a numerically exact treatment of nonequilibrium charge transport  in nanosystems with strong electronic-vibrational coupling. It covers a broad spectrum of parameters ranging from the nonadiabatic to the adiabatic regime and including both resonant and off-resonant transport. Being a nonperturbative method that includes all nonequilibrium effects, it allows a comprehensive description of this complex transport problem, as demonstrated here, for example, in the analysis of the structures and line shapes of the IETS. In the current formulation, the use of the exponential expansion of the bath correlation functions limits the application to moderate and high temperatures. Recent proposals\cite{Tian2012,Tang2015} to overcome this limitation appear promising. The implementation of such improved schemes as well as the extension of the method to describe current fluctuations will be the subject of future work.


We thank P.B. Coto for fruitful and inspiring discussions. This work was supported by the German Research Foundation (DFG). Generous allocation of computing time at the computing
center Erlangen (RRZE) is gratefully acknowledged.





%
\clearpage
\bibliography{./mybib}

\end{document}